\documentclass[journal]{IEEEtran}

\usepackage{amsmath,amssymb,amsfonts}
\usepackage{soul}
\usepackage{color}
\usepackage{url}
\usepackage{cite}
\usepackage[justification=centering]{caption}
\usepackage[detect-all]{siunitx}
\usepackage{graphicx}
\usepackage{pgfplots}
\usepackage{tikzscale}
\usepackage[siunitx]{circuitikz}
\usetikzlibrary{shapes.multipart}
\ifCLASSOPTIONcompsoc
\usepackage[caption=false,font=normalsize,labelfont=sf,textfont=sf]{subfig}
\else
\usepackage[caption=false,font=footnotesize]{subfig}
\fi

\usetikzlibrary{external}
\usepgfplotslibrary{external} 
\usepgfmodule{nonlineartransformations}
\usepackage{datatool}

\newcommand{\s}[1]{\mathrm{#1}}
\newcommand{\ph}[3][]{\varphi_\s{#2}^\s{#1}(#3)}  
\renewcommand{\j}{\s{j}}

\let\originalleft\left
\let\originalright\right
\renewcommand{\left}{\mathopen{}\mathclose\bgroup\originalleft}
\renewcommand{\right}{\aftergroup\egroup\originalright}

\usepackage{fancyhdr}
\fancypagestyle{empty}{fancy}

\begin{document}

\title{Arbitrary Angle of Arrival in\\Radar Target Simulation}
%
%
%

\author{Axel~Diewald,~\IEEEmembership{Graduate Student Member,~IEEE,}
        Benjamin~Nuss,~\IEEEmembership{Graduate Student Member,~IEEE}
        Mario~Pauli,~\IEEEmembership{Member,~IEEE}
        and~Thomas~Zwick,~\IEEEmembership{Fellow,~IEEE}
\thanks{Manuscript received Month Day, 2021; revised Month Day, 2021. This work was supported in part by the German Federal Ministry for Economic Affairs and Energy (BMWi) under Grant ZF4734201PO9 \textit{(Corresponding author: Axel Diewald.)}}%
\thanks{A. Diewald, B. Nuss, M. Pauli and T. Zwick are with the Institute of Radio Frequency Engineering and Electronics (IHE), Karlsruhe Institute of Technology (KIT), 76131 Karlsruhe, Germany (e-mail: axel.diewald@kit.edu).}
\thanks{Color versions of one or more of the figures in this article are available online at http://ieeexplore.ieee.org.}%
\thanks{Digital Object Identifier ...}%
}

\maketitle

\begin{abstract}
Automotive radar sensors play a key role in the current development of autonomous driving. Their ability to detect objects even under adverse conditions makes them indispensable for environment-sensing tasks in autonomous vehicles. The thorough and in-place validation of radar sensors demands for an integrative test system. Radar target simulators (RTS) are capable of performing over-the-air validation tests by creating artificial radar echoes that are perceived as targets by the radar under test. Since the authenticity and credibility of these targets is based on the accuracy with which they are generated, their simulated position must be arbitrarily adjustable. In this paper, a new approach to synthesize virtual radar targets at an arbitrary angle of arrival is presented. The concept is based on the superposition of the returning signals of two adjacent RTS channels. A theoretical model describing the basic principle and its constraints is developed. Additionally, a measurement campaign is conducted that verifies the practical functionality of the proposed scheme.


\end{abstract}


%
\IEEEpeerreviewmaketitle

\section{Introduction}

\IEEEPARstart{I}{n recent} years the development of advanced driver assistance systems (ADAS) and autonomous driving has reached new levels of sophistication. For the task of sensing the surrounding environment autonomous vehicles rely on a variety of sensors, such as camera, lidar (light detection and ranging), ultrasound and radar. Due to its weather robustness and long range capability, the latter plays a significant role for a large share of autonomous driving functions and therefore needs to be thoroughly and integratively validated. Carrying out these validation tests in the field involves a great deal of effort, as distances in the order of several million kilometers have to be covered to guarantee the faultless functioning of the system \cite{MGLW2015,S2017,KW2016}. In addition, these tests are not repeatable since individual traffic situations are unique and, therefore, must be reiterated whenever the system undergoes any design changes.

For these reasons, radar target simulators (RTS) have recently drawn a lot of attention in research, as they provide validation capabilities to test radar sensors in-place and under laboratory conditions \cite{GSGVABMPP2018, IRW2020, WMNLD2020}. Their working principle is to deceive a radar under test (RuT) by creating an artificial environment comprising of virtual radar targets. In order for this environment to be as credible and realistic as possible, the virtual radar targets must be generated as accurate as possible in regards of their characteristics. Recent RTS systems have already achieved a setting point precision higher than the resolution of common and even future radars in terms of range, Doppler and radar cross section (RCS) \cite{GMS18, SD2018, 9371304, 9448793}. Nonetheless, the simulation of the angle of arrival (AoA) in current RTS systems has yet to meet the angle estimation capabilities of their counterpart. By electronically switching between discrete and fixed angular positions the azimuth dislocation of a virtual radar target can be simulated \cite{GMS18,RS2021}, which, however, does not satisfy the angular accuracy capabilities of modern radar sensors. Another approach is to mechanically rotate the RTS system centric around the RuT \cite{9337461,GR2017,KC20xx}, which significantly limits the number of virtual targets and their inherent lateral movement speed. Rotating the RuT itself \cite{KT2017} results in the same restrictions and, in addition, is not suitable for integrated radar sensor validation.


Therefore, the authors present a new approach that enables the generation of virtual radar targets with an arbitrary angle of arrival that is neither limited in regards of the methodology of the RTS, as it is applicable for analog and digital systems, nor by the modulation scheme of the RuT. The concept is based on the superposition of two neighboring virtual radar echoes, that enables the synthesis of simulated radar targets at an adjustable lateral position. In the following the working principle of radar target simulation and the underlying signal model will be outlined. Thereupon, the fundamental idea of the proposed approach, as well as its constraints, calibration and disadvantages will be elaborated. Finally, the results of a measurement campaign that demonstrate the successful implementation of the concept will be presented.


\section{Radar Target Simulation}

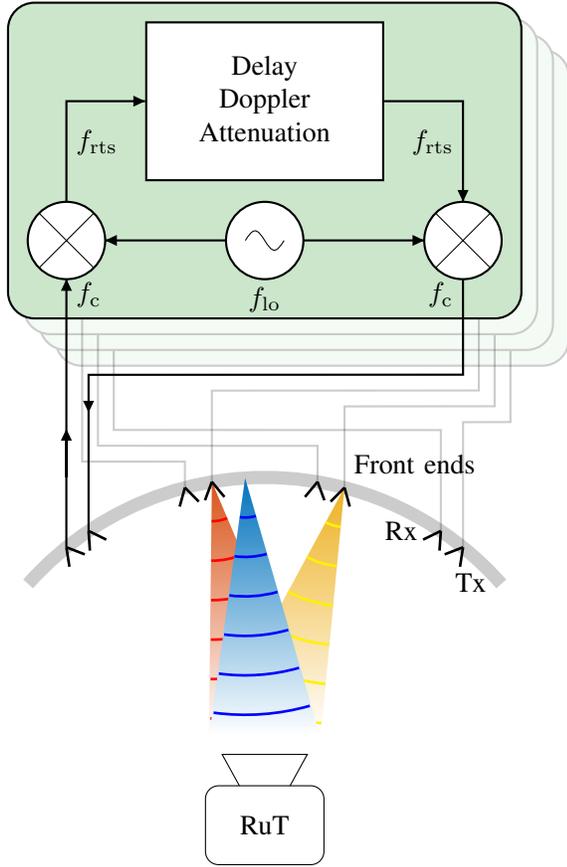
\begin{figure}
	\centering
	\definecolor{mygrey}{rgb}{.8,.8,.8}
\definecolor{mygreen}{rgb}{.8,.9,.8}
\definecolor{myblue2}{rgb}{0.00000,0.44700,0.74100}%
\definecolor{myred}{rgb}{0.85000,0.32500,0.09800}%
\definecolor{myyellow}{rgb}{0.92900,0.69400,0.12500}%

\ctikzset{
	mylength/.style={bipoles/length=#1}
}

\def\myscale{1.05}

\def\Nfe{4}
\def\Nfem{3}
\pgfmathsetmacro{\ferad}{4}
\pgfmathsetmacro{\fea}{48.5}
\pgfmathsetmacro{\fesegang}{2*\fea/\Nfe}

\pgfmathsetmacro{\boxheight}{4}

\begin{circuitikz}[scale=\myscale, every node/.style={scale=\myscale},>=latex]

	\draw
	(0,0) node[] (RuT) {}
	node[ampshape,line width=0.2pt,fill=white,color=black, rotate=270] {}
	++(0,.1)
	node[draw,line width=0.4pt,minimum width=1.5cm,minimum height=1cm,anchor=north,fill=white,rounded corners=5] () {RuT}
	
	;
	
	
	\draw [line width=5pt,color=mygrey] (RuT) ++(90+\fea:\ferad) arc [start angle=90+\fea,delta angle=-\fea*2,radius=\ferad];
	
	\foreach \i in {0,...,\Nfem}{
		\draw (RuT) ++({90+\fea-\fesegang*(\i+.5)+2.5}:\ferad) node[](ferx\i){};
		\draw (RuT) ++({90+\fea-\fesegang*(\i+.5)-2.5}:\ferad) node[](fetx\i){};
		\coordinate (ferxc\i) at (ferx\i);
		\coordinate (fetxc\i) at (fetx\i);
	}
	
	
	
	\shade[top color=myred] (fetxc1)--({90+45}:1)--({90-45}:1)--cycle;
	\foreach \i in {0,...,5}{
		\draw [line width=1pt,color=red] (fetxc1)++({180+(90+\fea-\fesegang*(1.5)-2.5)-10}:{\ferad-1-\i/2}) arc [start angle={180+(90+\fea-\fesegang*(1.5)-2.5)-10},delta angle=23,radius={\ferad-1-\i/2}];
	}
	
	\shade[top color=myyellow] (fetxc2)--({90+45}:1)--({90-45}:1)--cycle;
	\foreach \i in {0,...,5}{
		\draw [line width=1pt,color=yellow] (fetxc2)++({180+(90+\fea-\fesegang*(2.5)-2.5)-14}:{\ferad-1-\i/2}) arc [start angle={180+(90+\fea-\fesegang*(2.5)-2.5)-14},delta angle=23,radius={\ferad-1-\i/2}];
	}
	\shade[top color=myblue2] (RuT) ++({90+\fesegang*(.25)-2.5}:\ferad)--({90+45}:1)--({90-45}:1)--cycle;
	\foreach \i in {0,...,5}{
		\draw [line width=1pt,color=blue] (RuT) ++({90+\fesegang*(.25)-2.5}:\ferad)++({180+90+\fesegang*(.25)-2.5-11.2}:{\ferad-1-\i/2}) arc [start angle={180+90+\fesegang*(.25)-11.2-2.5},delta angle=23.5,radius={\ferad-1-\i/2}];
	}
	
	
	
	\foreach \i in {0,...,\Nfem}{
		\draw [line width=1pt] (ferxc\i) -- ++({270+\fea-\fesegang*(\i+.5)+2.5-30}:0.25);
		\draw [line width=1pt] (ferxc\i) -- ++({270+\fea-\fesegang*(\i+.5)+2.5+30}:0.25);
	}
	\foreach \i in {0,...,\Nfem}{
		\draw [line width=1pt] (fetxc\i) -- ++({270+\fea-\fesegang*(\i+.5)-2.5-30}:0.25);
		\draw [line width=1pt] (fetxc\i) -- ++({270+\fea-\fesegang*(\i+.5)-2.5+30}:0.25);
	}
	
	\draw (RuT) ++(0,6) node[] (RTS) {};
	
	\foreach \i in {\Nfem,...,1}{
		\draw (RTS)++(0.2*\i,-0.2*\i) node[draw,thick,minimum width=6.5cm,minimum height=\boxheight cm,rounded corners=10,fill=white,white,anchor=south] {};
		\draw (RTS)++(0.2*\i,-0.2*\i) node[draw,thick,minimum width=6.5cm,minimum height=\boxheight cm,rounded corners=10,fill=mygreen, opacity=0.2,anchor=south] {};
	}
	
	\draw (RTS)
	node[draw,thick,minimum width=6.5cm,minimum height=\boxheight cm,rounded corners=10,fill=mygreen,anchor=south] (rts_box) {}
	
	(rts_box.south)  ++(-3,1)
	node[mixer, anchor=west,fill=white] (mix_lo_rx) {}
	(mix_lo_rx.south) node[inputarrow,rotate=90] {}
	(mix_lo_rx.east) node[inputarrow,rotate=180] {}
	
	(rts_box.south)  ++(3,1)
	node[mixer, anchor=east,fill=white] (mix_lo_tx) {}
	(mix_lo_tx.north) node[inputarrow,rotate=270] {} 
	(mix_lo_tx.west) node[inputarrow,rotate=0] {}
	
	(rts_box.south) ++(0,1) node[vcoshape,fill=white] (vco) {}  ++(0,-.75) node[]{$f_\mathrm{lo}$};
	\draw [line width=0.8pt] (vco.west) -- (mix_lo_rx.east)
	(vco.east) -- (mix_lo_tx.west); 
	
	\draw (rts_box.south) ++(0,1.75) node[draw,line width=0.8pt,minimum width=3cm,minimum height=2cm,anchor=south,fill=white] (target_gen) {\begin{tabular}{c} Delay \\ Doppler \\ Attenuation \end{tabular}};
	\draw (target_gen.west) node[inputarrow,rotate=0] {};
	
	\draw [line width=0.8pt] (mix_lo_rx.north) -- (mix_lo_rx.north |- target_gen.west)  -- (target_gen.west);
	\draw [line width=0.8pt] (target_gen.east) -- (target_gen.east -| mix_lo_tx.north)  -- (mix_lo_tx.north);
	

	
	\foreach \i in {0,...,\Nfem}{
		\draw (mix_lo_rx.south |- 0,4+\i*.2)++(\i*.2,0) node[] (feirx1\i) {};
		\draw (mix_lo_tx.south |- 0,5.3-\i*.2)++(\i*.2,0) node[] (feitx1\i) {};
		\coordinate (ferxp2\i) at (feirx1\i);
		\coordinate (fetxp2\i) at (feitx1\i);
	}
	
	\draw [line width=0.8pt] (ferxc0) -- (ferxc0 |- 0,4.5) -- (ferxp20) -- (mix_lo_rx.south);
	\draw [line width=0.8pt] (fetxc0) -- (fetxc0 |- 0,5.3) -- (fetxp20) -- (mix_lo_tx.south);
	\foreach \i in {1,...,\Nfem}{
		\draw [opacity=0.2,line width=0.8pt] (ferxc\i) -- (ferxc\i |- 0,4+\i*.2) -- (ferxp2\i) -- ++(0,2.2-\i*.4);
		\draw [opacity=0.2,line width=0.8pt] (fetxc\i) -- (fetxc\i |- 0,5.3-\i*.2) -- (fetxp2\i) -- ++(0,.9);
	}
	
	\draw (ferxc0)++(0,1.5) node[inputarrow,rotate=90] {};
	\draw (fetxc0)++(0,1.5) node[inputarrow,rotate=270] {};
	
	

	
	\draw (ferx3)++(-.5,0)  node[align=left] {Rx};
	\draw (fetx3)++(.1,-.45)  node[align=left] {Tx};
	
	\draw (fetx2)++(0,.3)  node[align=left,anchor=west] {Front ends};
	
	\draw (mix_lo_rx.south) ++(0,-.2) node[align=left,anchor=west] {$f_\mathrm{c}$};
	\draw (mix_lo_tx.south) ++(0,-.2) node[align=right,anchor=east] {$f_\mathrm{c}$};
	
	\draw (mix_lo_rx.north) ++(0,.75) node[align=left,anchor=west] {$f_\mathrm{rts}$};
	\draw (mix_lo_tx.north) ++(0,.75) node[align=right,anchor=east] {$f_\mathrm{rts}$};

\end{circuitikz}
	\caption{Concept of the radar target simulator}
	\label{fig_rts_concept}
\end{figure}

The overall concept of the RTS is shown in Fig. \ref{fig_rts_concept}. The RuT is placed closely in front of the RTS antenna front ends in compliance with the far-field condition \cite{longhurst1986geometrical}. The front end modules are arranged in a semicircle formation with the RuT as the center and equal distance between each module. The receive antenna (Rx) picks up the radar signal transmitted by the RuT, which is thereafter down converted to a lower intermediate frequency $f_\s{rts}$. Subsequently, the single target generation modifications, namely a time delay, a Doppler shift and an attenuation, are applied to the signal before it is up converted back to its original carrier frequency $f_\mathrm{c}$ and re-transmitted towards the RuT by the front end transmit antenna (Tx). Each RTS front end pair is coupled with its own signal modification module enabling an independent target generation for each RTS channel. 

\subsection{Target Generation}

RTS systems can be divided into analog or digital in terms of their target generation methodology. Since the approach presented in this paper can be implemented with either one, both system domains are shortly explained.

Analog RTS systems utilize optical or electrical delay lines \cite{EPB2016,LEWW2014,GGSABMP2017}, surface acoustic wave (SAW) filters \cite{ALPTB2017}, or frequency mixers \cite{RH2020,IRW2020} to simulate the target's range. The required Doppler shift can be applied with a vector modulator \cite{GSGVABMPP2018,IMSW2019}, with a digitally controlled phase shifter \cite{OZH2019}, or by applying a frequency offset to the local oscillator for the up and down conversion \cite{EPB2016}. A variable gain amplifier can be employed for the simulation of the RCS.

Digital RTS systems, on the other hand, perform the target generation task utilizing a field-programmable gate array (FPGA) after a preceding analog-to-digital conversion. Within the FPGA the radar signal is either analyzed and re-synthesized \cite{WMNLD2020}, or modified and looped back in order to create the respective virtual targets. The modification regarding the delay can be realized by sample buffering \cite{SN2020,ZZGS2020} or with a digital finite impulse response (FIR) filter \cite{SD2018}. The velocity can be simulated with the help of a complex quadrature mixer \cite{SN2020,ZZGS2020} or through fine range discretization \cite{SMD2019}. The attenuation can be implemented with a simple samplewise multiplication.

The delay $\tau_\s{rts}$, Doppler shift $f_\s{D,rts}$ and attenuation $A$ required for the virtual target generation can be derived from the target's range $R_\s{t}$, radial velocity $v_\s{t}$ and RCS $\sigma_\s{t}$ as follows
\begin{align}
	\tau_\s{rts} &= \frac{2R_\s{t}}{c_\mathrm{0}} \\
	f_\mathrm{D,rts} &= \frac{2f_\mathrm{c}v_\s{t}}{c_\mathrm{0}} \\
	A &= \frac{\sqrt{\sigma_\s{t}}}{{R_\s{t}}^2}
\end{align}
where $f_\s{c}$ describes the lower bound of the radar's frequency band and $c_\s{0}$ the speed of light.

\subsection{Signal Model}
In the following the radar signal that is transmitted by the RuT and modified by the RTS will be modeled. For the sake of simplicity, a Frequency-Modulated Continuous Wave (FMCW) radar will be assumed. Nonetheless, the underlying principle of this approach operates independently of the modulation scheme of the radar and only the subsequent mathematical expressions must be adapted. The RuT's multiple-input multiple-output (MIMO) antenna array, comprising of $N_\s{tx}$ transmit and $N_\s{rx}$ receive antenna elements, can be unified to form a virtual antenna array of size $N_\s{A} = N_\s{tx} \cdot N_\s{rx}$ \cite{1291865}. As will later be shown, the signal delay's impact on the signal phase, plays a key role for the success of the proposed concept. For this reason, the following analytical descriptions focus primarily on the signal phase in order to facilitate the comprehension of the approach and its limitations.

The RuT transmits a signal whose frequency and phase can be described in regards of time $t \in [0,T]$ as
\begin{align}
	f_\s{tx}(t) &= f_\s{c} + \frac{B}{T} \cdot t \\
	\ph{tx}{t} &= 2\pi \int_0^{t} f(t')\cdot dt' = 2\pi \left[f_\s{c} \cdot t + \frac{B}{2T} \cdot t^2\right]
\end{align}
where $B$ is the signal's bandwidth and $T$ the chirp period. The radiated signal travels through free space, is received by one of the RTS front ends and down converted, causing a time delay $\tau_\s{tx}$ and a phase shift of $\ph[]{lo}{t} = 2\pi f_\s{lo} \cdot t$ and can be expressed as
\begin{align}
	\ph{rts,rx}{t}&= \ph{tx}{t-\tau_\s{tx}} - \ph[]{lo}{t} \nonumber \\ 
	&= 2\pi \left[-f_\s{c} \tau_\s{tx} + f_\s{rts} \cdot t + \frac{B}{2T}  (t-\tau_\s{tx})^2\right]
\end{align}
where $f_\s{rts} = f_\s{c} - f_\s{lo}$ describes the signal's down converted lower bound frequency. The RTS generates a virtual radar target by applying an artificial delay $\tau_\s{rts}$. Subsequently, the signal is up converted to its carrier frequency
\begin{align}
	\ph{rts,tx}{t} =& \ \ph{rts,rx}{t-\tau_\s{rts}} + \ph[]{lo}{t} \nonumber \\
	=& \  2\pi \bigg[-f_\s{c} \tau_\s{tx} - f_\s{rts} \tau_\s{rts} + f_\s{c} \cdot t  \nonumber \\
	&\qquad\!\! + \left. \frac{B}{2T} (t-\tau_\s{tx}-\tau_\s{rts})^2\right]
\end{align}
For simplicity, no Doppler shift is applied and therefore the Doppler estimation will later be skipped without compromising the generality of the approach, as will be shown later. The signal is re-transmitted and received by the RuT causing a further delay $\tau_\s{rx}$. Thereupon, it is mixed with the transmit signal to form the complex beat signal
\begin{align}
	\ph{b}{t} =& \ph{tx}{t} - \ph{rts,tx}{t-\tau_\s{rx}} \nonumber \\
	=& \ 2\pi \left[f_\s{c} \tau_\s{c} + f_\s{rts} \tau_\s{rts} + \frac{B}{2T} (2\tau \cdot t -{\tau}^2)\right]
\end{align}
where $\tau_\s{c} = \tau_\s{tx} + \tau_\s{rx}$ denotes the free space propagation delay and $\tau = \tau_\s{c} + \tau_\s{rts}$ the total signal delay. Following this, the radar signal is discretized $(t = n_\s{s}/f_\s{s})$ by the RuT's analog-to-digital converter (ADC)
\begin{align}
	x_\s{b}[n_\s{s}] = A \cdot \exp \left\{ \j \varphi_\s{b}\left(\frac{n_\s{s}}{f_\s{s}}\right) \right\}
\end{align}
where $n_\s{s} \in [0,N_\s{s}-1]$ represents the sample index, $f_\s{s}$ the sampling frequency and $A$ the signal amplitude. A discrete Fourier transform (DFT) is applied to estimate the target's range
\begin{align}
	x_\s{R}[f_\s{R}] =& \sum_{n_\s{s}=0}^{N_\s{s}-1} x_\s{b}[n_\s{s}] \cdot \exp \left\{ -\j2\pi\frac{n_\s{s}f_\s{R}}{N_\s{s}} \right\}
\end{align}
where $f_\s{R} \in [0,N_\s{s}-1]$ designates the DFT bin index. The expression can be simplified using the partial sum of a geometric series \cite{Bron01} and $\sin(x) \approx x$ for $|x| \ll 1$ to
\begin{align}
	x_\s{R}[f_\s{R}] =& \ A N_\s{s} \cdot \s{sinc} \left( \frac{B}{N_\s{s}} \tau - \frac{f_\s{R}}{N_\s{s}} \right) \cdot \exp \left\{ \j \varphi_\s{R}[f_\s{R}] \right\} \label{eq_x_range} \\
	\varphi_\s{R}[f_\s{R}] =& 2\pi \left[f_\s{c} \tau_\s{c} + f_\s{rts} \tau_\s{rts} + \frac{1}{2} \left( B \tau -  f_\s{R} \right)\right]
\end{align}
The $\s{sinc}$-function reaches its maximum where its argument equals zero, leading to a target being detected at $f_\s{R} = B\tau$. Considering the RuT's virtual antenna array with an antenna spacing of $d = \lambda/2$, the returning signal's delay
\begin{align}
	\tau_\s{rx} = \frac{R_\s{c}+d \sin(\theta) \cdot n_\s{A}}{c_\s{0}}
\end{align}
is dependent on the angle of arrival (AoA) $\theta \in [\SI{-90}{\degree},\SI{90}{\degree}]$, which describes the incidence angle in the azimuth plane and is centered in the direction of the RTS. $n_\s{A} \in [0,N_\s{A}-1]$ denotes the antenna element index and $R_\s{c}$ the physical distance between the antenna element and the RTS and was set equal for all elements for simplicity. By applying beamforming the AoA of a detected target can be estimated
\begin{align}
	x_\s{A}[\alpha] =& \sum_{n_\s{A}=- \frac{N_\s{A} - 1}{2}}^{\frac{N_\s{A} - 1}{2}} x_\s{R}[f_\s{R}] \cdot \exp \left\{ \j\pi\sin(\alpha) \cdot n_\s{A} \right\}
\end{align}
where $\alpha \in [\SI{-90}{\degree},\SI{90}{\degree}]$ is orientated equal to $\theta$ and can be discretized arbitrarily. Simplifying in the same manner as before, the expression results in
\begin{align}
	x_\s{A}[\alpha] =& \ A N_\s{s} N_\s{A} \cdot \s{sinc} \left( \frac{\sin(\theta)-\sin(\alpha)}{2} \right) \cdot \exp \left\{ \j \varphi_\s{A}  \right\} \label{eq_x_fe} \\ 
	\varphi_\s{A} =& \ 2\pi \left[ \left( f_\s{c} + \frac{B}{2} \right) \frac{2 R_\s{c}}{c_\s{0}} + \left( f_\s{rts} + \frac{B}{2} \right) \tau_\s{rts} \right. \nonumber \\
	& \left. \qquad + \sin(\theta) \frac{N_\s{A} - 1}{4} \right] \label{eq_ph_fe}
\end{align}
Again, the maximum of the $\s{sinc}$-function yields the detected AoA at $\sin(\alpha) = \sin(\theta)$.


\section{Arbitrary Angle of Arrival}\label{sec_aaoa}

The principle of the approach lies in the superimposing of two radar signals returning from adjacent RTS channels to form a composed virtual radar target whose AoA resides between the physical angular positions of the respective RTS front ends and can be adjusted arbitrarily. The alteration is performed through the signal attenuation within the target generation process. In the following, the resulting superimposed signal and the required RTS channel amplitude adaptations are derived analytically. Furthermore, the constraints for a successful arbitrary AoA simulation and the necessary calibration to meet them are developed and the arising disadvantages are concluded.

\subsection{Superposition of target signals}
\begin{figure}[!t]
\centering
\includegraphics{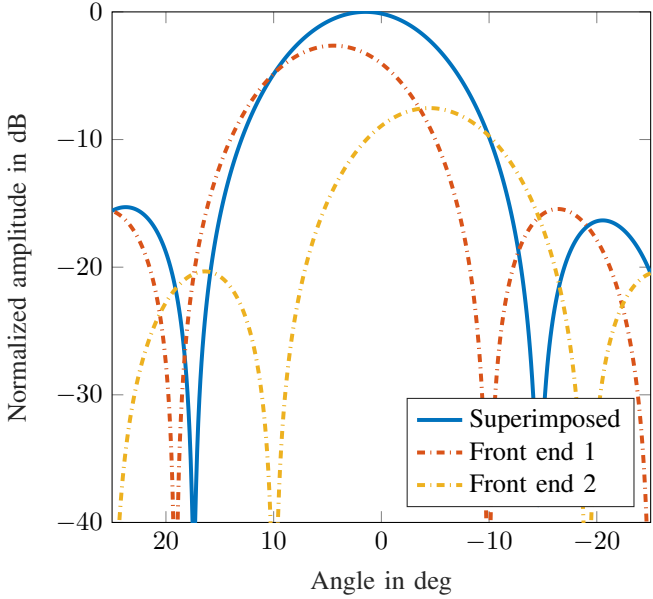}
\caption{Simulated signal superposition of two adjacent RTS channels after beamforming}
\label{fig_sim_superpos}
\end{figure}

The radar signals returning from the RTS front ends can be considered as free of mutual interference, since the electromagnetic waves do not interfere in free space and only superimpose at the antenna element. Therefore, the signals only dependent on the emitting RTS channel and its respective, relative front end angle $\theta_\s{q}$. Setting forth \eqref{eq_x_fe} and \eqref{eq_ph_fe}, they can be expressed as
\begin{align}
	x_\s{A,q}[\alpha] =& \ A_\s{q} N_\s{s} N_\s{A} \cdot \s{sinc} \left( \frac{\sin(\theta_\s{q})-\sin(\alpha)}{2} \right) \nonumber \\
	&\cdot \exp \left\{ \j \varphi_\s{A,q}  \right\} \\
	\varphi_\s{A,q} =& \ 2\pi \left[ \left( f_\s{c} + \frac{B}{2} \right) \frac{2 R_\s{c,q}}{c_\s{0}} + \left( f_\s{rts} + \frac{B}{2} \right) \tau_\s{rts,q} \right. \nonumber \\
	& \left. \qquad + \sin(\theta_\s{q}) \frac{N_\s{A} - 1}{4} \right] \label{eq_ph_fe_q}
\end{align}
where $q \in [1,2]$ indexes two neighboring RTS channels. The complex-valued superposition of these is graphically depicted in Fig. \ref{fig_sim_superpos} and can be described as
\begin{align}
	\widehat{x}_\s{A}[\alpha] =& \ x_\s{A,1}[\alpha] + x_\s{A,2}[\alpha] \label{eq_x_sup} 
\end{align}

In order to determine how the relation of the RTS channels' attenuation controls the maximum of the superimposed signal and therefore the detected AoA, the previous expression is derived according to $\alpha$ and set equal to zero leading to
\begin{align}
	\frac{\partial \widehat{x}_\s{A}[\alpha]}{\partial \alpha} &= \frac{\partial x_\s{A,1}[\alpha]}{\partial \alpha} + \frac{\partial x_\s{A,2}[\alpha]}{\partial \alpha} = 0 \label{eq_superpos_deriv}
\end{align}
The derivation of the individual RTS channel signals can be expressed as
\begin{align}
	\frac{\partial x_\s{A,q}[\alpha]}{\partial \alpha} = A_\s{q} N_\s{s} N_\s{A} \cdot g_\s{q}(\alpha) \cdot \exp \left\{ \j \varphi_\s{A,q}  \right\} 
\end{align}
with the substitute for the derivative of the $\s{sinc}$-function 
\begin{align}
	g_\s{q}(\alpha) =& \frac{2 \cdot \cos \left( \frac{\sin(\theta_{q}) - \sin(\alpha)}{2} \right)}{\sin(\theta_{q}) - \sin(\alpha) } \nonumber \\
	& - \frac{4 \cdot \sin \left( \frac{\sin(\theta_{q}) - \sin(\alpha)}{2} \right)}{ \left( \sin(\theta_{q}) - \sin(\alpha) \right)^2 }
\end{align}
Solving \eqref{eq_superpos_deriv} for the amplitude relation results in
\begin{align}
	\frac{A_\s{1}}{A_\s{2}} = \exp \left\{ \j \left(\varphi_\s{\alpha,1}-\varphi_\s{\alpha,2}\right) \right\} \cdot \frac{g_\s{2}(\alpha)}{g_\s{1}(\alpha)} \label{eq_amp_rel}
\end{align}
which can be utilized to calculate and set the required attenuation for a specified target AoA.

\subsection{Constraints}
The constructive interference that is needed in order to synthesize and steer the arbitrary AoA is subject to certain constraints. First, the individual signals' respective radar targets must be detected in the same range and velocity bin, so that they will be superimposed in the succeeding beamforming processing. Next, the spacing of the adjacent front ends has to be less than or equal to the RuT's angular resolution $\Delta \alpha$ \cite{hecht2012optics}
\begin{align}
	\Delta\theta &= \theta_\s{1} - \theta_\s{2} \leq 1.22 \frac{\lambda}{d_\s{A}} = \Delta \alpha
\end{align}
where $d_\s{A}$ is the size of the aperture of the RuT's virtual antenna array, otherwise the composite signal will form two individual peaks instead of a common one, which will be detected unintentionally and potentially as individual targets. Finally, the presented approach depends on the phase coherence of the individual signals so that they overlap in a purely constructive manner, or else the signals will partially or completely cancel each other out, resulting in a distorted angle detection. For this reason, it is important to ensure phase coherency. For a constructive interference the phase controlling terms in \eqref{eq_x_sup} must be set equal,
\begin{align}
	\exp \left\{ \j \varphi_\s{A,1}  \right\} = \exp \left\{ \j \varphi_\s{A,2}  \right\},
\end{align}
from which the following expression can be derived
\begin{align}
	\Delta\varphi &= \varphi_\s{A,1} - \varphi_\s{A,2} = 2 k \pi, \qquad k \in \mathbb{N} \label{eq_constr_ph}
\end{align}
As it can be concluded from \eqref{eq_ph_fe_q}, the phase of the individual signals is very sensitive to deviations of the physical distance between the RuT and the front ends. Even the smallest inaccuracies in the mechanical mounting of the front ends lead to a significant relative phase shift and potentially the extinction of the composite signal. For a relative radial position difference of $\lambda / 4 \approx \SI{973}{\micro\meter}$, assuming a carrier frequency of $f_\s{c} = \SI{77}{\giga\hertz}$, the phase difference causes destructive interference. Luckily, the phase also lies under the influence of the simulative delay $\tau_\s{rts}$ which can be adjusted independently for the individual RTS channels and therefore be utilized to establish the required phase coherency.

\subsection{Calibration}
Intuitively, achieving phase coherency would necessitate a precise estimation of the respective phases prior to their adjustment. However, as has been shown in \cite{SWB1010788574}, high-precision radar range estimation down to the order of fractions of the wavelength, as it is required in this case, demands for a high signal-to-noise ratio (SNR), and can nevertheless only provide a relative rough estimate of the absolute phase together with a statistical specification for its estimation accuracy.

Therefore, the authors propose an alternative solution for the calibration of the respective RTS channel phases. The difference between the set-point and the actual value of the target angle (angle error $\alpha_\s{\epsilon}$) stays in relation to the relative phase offset \eqref{eq_constr_ph} between the respective RTS channels. The angle error reaches minima when the relative phase offset equals multiples of $2\pi$. Therefore, the calibration can be executed in two steps. First, the target ranges must be calibrated to match within a range bin and subsequently a fine parameter sweep for $\tau_\s{rts}$ is performed in order to find the the minimum for $\alpha_\s{\epsilon}$. Deriving \eqref{eq_ph_fe_q} according to $\tau_\s{rts}$ reveals the influential factor of a simulative delay offset $\Delta \tau_\s{rts}$ on the signal's phase shift 
\begin{align}
	\frac{\partial \varphi_\s{A,q}}{\partial \tau_\s{rts}} =& \ 2\pi \left( f_\s{rts} + \frac{B}{2} \right) \label{eq_infl_tau}
\end{align}
which can be utilized to specify the span and step size of the parameter sweep. The calibration process should cover at least one complete rotation of the phase ($2\pi$), which determines the interval of $\Delta\tau_\s{rts}$ for the sweep.

In addition to the phase, the amplitude must also be calibrated, since its deviations directly translate into an angle error. This can be accomplished relatively easily by measuring the power of front ends individually, during which only one of them is active at a time.

\subsection{Disadvantages}
On the downside of the presented calibration method, the superimposed $\s{sinc}$-functions in the range domain \eqref{eq_x_range} experience a small range offset due to $\tau_\s{c,1} + \tau_\s{rts,1} \neq \tau_\s{c,2} + \tau_\s{rts,2}$, broadening the peak and thereby diminishing the achievable range accuracy of the measurement. This offset is caused by the differing influences that $\tau_\s{c}$ and $\tau_\s{rts}$ have on the phase, which depend on the respective signal frequency ($f_\s{c}$ and $f_\s{rts}$) at which they are applied. This peculiarity is exploited in calibration method shown here, but at the same time results in a range offset, since the total delay $\tau$ is a simple addition of the two aforementioned.

Furthermore, since the receive antennas of the RTS are localized in the immediate vicinity of the RuT, the far-field condition \cite{longhurst1986geometrical} is not met, and the virtual array assumption is not fully accurate. For this reason, the incoming signal phases at the individual RTS receivers exhibit a mutual offset which inevitably causes an error during the angle estimation.


\section{Measurement}

\begin{figure}[!t]
	\centering
	\input{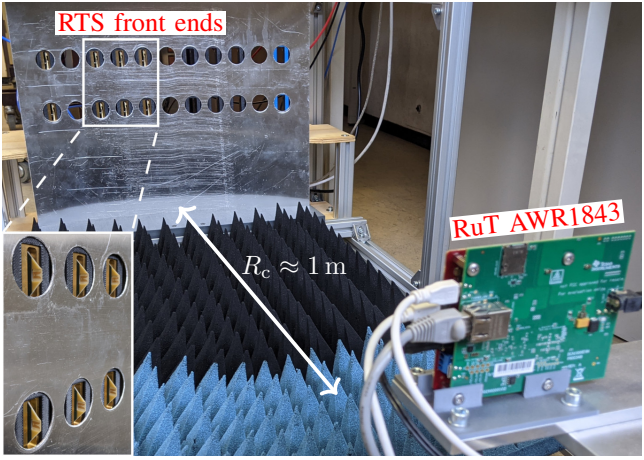}
	\caption{Measurement setup}
	\label{fig_meas_setup}
\end{figure}

\begin{figure}[!t]
	\centering
	\includegraphics{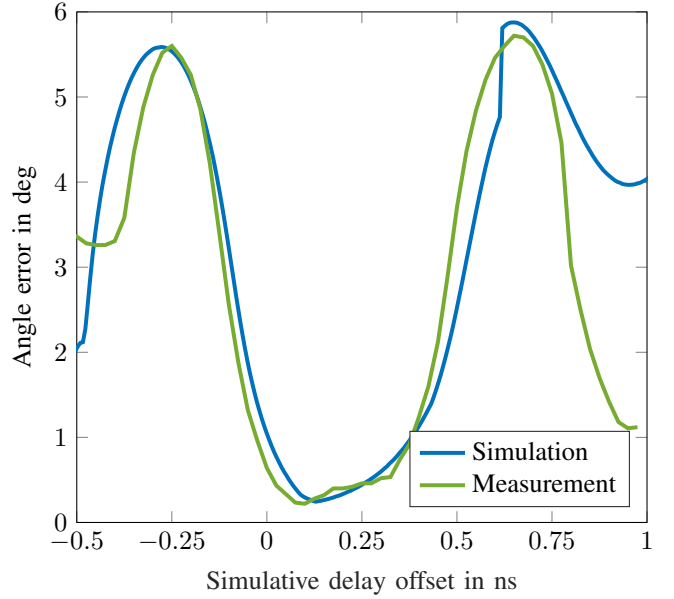}
	\caption{Phase coherency calibration}
	\label{fig_calib}
\end{figure}

For the measurement campaign, a digital RTS system, whose basic operational components are described in \cite{9448793} and \cite{vehicles3020016}, is utilized. The modular front ends were arranged in a semicircle formation at a distance of $R_\s{c} = \SI{1}{\meter}$ and behind a curved metal sheet with round cutouts for the front end antennas. The metal sheet served the purpose of facilitating the positioning of the front ends and the blockage of undesired radar reflections of the background, leaving only determinable static reflections. Fig.~\ref{fig_meas_setup} shows a photo of the measurement setup.

An UltraScale+ RFSoC FPGA from Xilinx was employed for the back end. The integrated ADCs and DACs were configured to a sampling rate of $f_\s{s} = \SI{4}{GSPS}$ and the intermediate frequency was set to $f_\s{rts} = \SI{500}{\mega\hertz}$. The application of the target's delay, Doppler shift and attenuation were realized through sample buffering, a complex quadrature mixer and a linear multiplier, respectively. However, sample buffering at the given sample rate only allows for an delay step size of $\Delta \tau_\s{rts,buf} = \SI{0.25}{\nano\second}$, which will later proof to be too coarse for the parameter sweep that is necessary for the calibration and achieving phase coherency. Therefore, fractional delay filters, that enable the application of arbitrary fractions of the sample period \cite{9448793}, were used.

As the RuT a Texas Instruments AWR1843BOOST radar sensor evaluation module with a bandwidth of $B = \SI{1}{\giga\hertz}$ and a lower bound frequency of $f_\s{c} = \SI{77}{\giga\hertz}$ was deployed. The radar features $N_\s{tx} = 2$ transmit and $N_\s{rx} = 4$ receive antennas with a spacing of the virtual antenna array of $\lambda/2$, leading to an angular resolution of $\Delta \alpha = \SI{17.5}{\degree}$.

%

\begin{figure}
	\centering
	\subfloat[Simulated amplitude attenuation\label{fig_ang_ampl}]{%
		\includegraphics{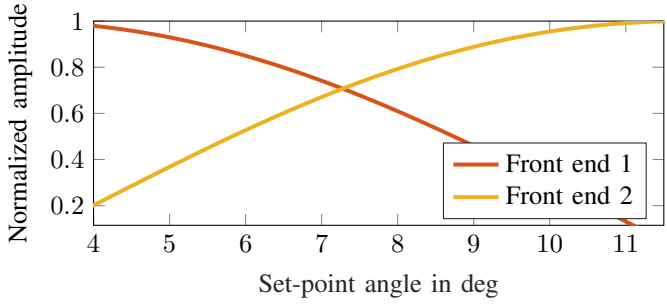}}\\
	\subfloat[Simulated and measured angle linearity\label{fig_ang}]{%
		\includegraphics{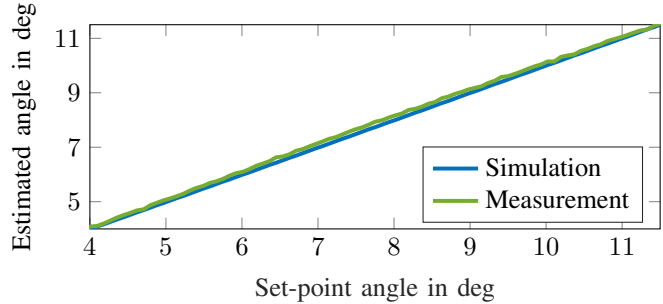}}\\
	\subfloat[Simulated and measured angle error\label{fig_ang_err}]{%
		\includegraphics{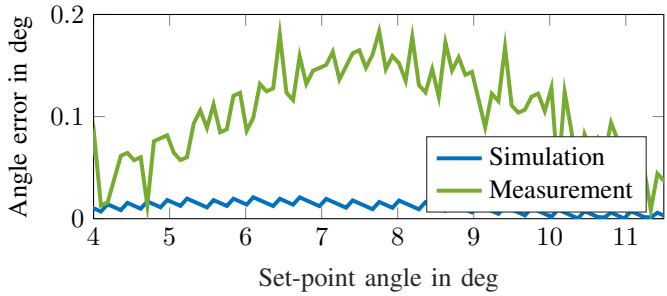}} 
	\caption{(a) Amplitude attenuation and angle~(b)~linearity~and~(c)~error}
	\label{meas_sim_ang_g} 
\end{figure}

For all measurements, two neighboring RTS channels with their front ends positioned at $\theta_\s{1} = \SI{3.4}{\degree}$ and $\theta_\s{2} = \SI{12.2}{\degree}$ were active. The calibration was performed by monitoring the angle error $\alpha_\s{\epsilon}$, keeping the delay of the first RTS channel $\tau_\s{rts,1}$ constant, while sweeping that of the second channel $\tau_\s{rts,2} = \tau_\s{rts,1} + \Delta \tau_\s{rts}$ in an interval of $\Delta \tau_\s{rts} = [\SI{-0.5}{\nano\second}, \SI{1}{\nano\second}]$ with a step size of $\SI{25}{\pico\second}$. Fig. \ref{fig_calib} depicts the simulated and measured angle error $\alpha_\s{\epsilon}$ resulting from the parameter sweep. As can be observed, the measured value arrives at a minimum for $\Delta \tau_\s{rts} = \SI{0.1}{\nano\second}$, which is used to calibrate the RTS channels for the subsequent measurements. The lateral distance between the two maxima is approximately $\SI{1}{\nano\second}$, which is consistent with \eqref{eq_infl_tau}, as it represents a phase shift of $\Delta \varphi = 2\pi$. The simulation was performed to verify the correctness of the preceding analytical derivations. The small difference between simulation and measurement can be explained by the inaccuracy, with which the physical positions of the front ends in regards of range and angle were determined by measurement, and to which the simulation was adapted.

Next, the achievable linearity and its corresponding error of the synthesized AoA was measured. For this, the required delay offset that was determined during the calibration was applied to the RTS channels in order to establish phase coherency. The angle set-point was linearly increased and the respective signal attenuations were determined according to \eqref{eq_amp_rel}. Fig. \ref{meas_sim_ang_g} displays the measured and simulated angle value, as well as its deviation from its set-points. Measurement and simulation both show good agreement between set and actual value, as the maximum angle error of $\SI{0.2}{\degree}$ correlates to only $\SI{1.14}{\percent}$ in relation to the angle resolution of the RuT. Furthermore, it can be assumed that the angle error occurs due to a remaining amplitude and phase offset between the utilized RTS channels. The latter could be further reduced by an iterative calibration with decreasing step sizes for $\Delta \tau_\s{rts}$, but can not be eliminated completely as the aforementioned restrictions still apply. The simulation again serves as a reference for the underlying theory developed in the previous chapter.

\begin{table}[!t]
	\renewcommand{\arraystretch}{1.3}
	\caption{Target Characteristics}
	\label{tab_targets}
	\centering
	\begin{tabular}{c|rrr}
		Target & Range & Velocity & Angle\\
		\hline
		1 & $\SI{33.5}{\meter}$ & $\SI[per-mode=repeated-symbol]{0}{\meter\per\second}$ & $\SI{7}{\degree}$\\
		2 & $\SI{37}{\meter}$ & $\SI[per-mode=repeated-symbol]{4}{\meter\per\second}$ & $\SI{4}{\degree}$\\
		3 & $\SI{45}{\meter}$ & $\SI[per-mode=repeated-symbol]{-2}{\meter\per\second}$ & $\SI{10}{\degree}$\\
		4 & $\SI{52}{\meter}$ & $\SI[per-mode=repeated-symbol]{-5}{\meter\per\second}$ & $\SI{11}{\degree}$\\
	\end{tabular}
\end{table}

\begin{figure}[!t]
	\centering
	\includegraphics[width=1\columnwidth]{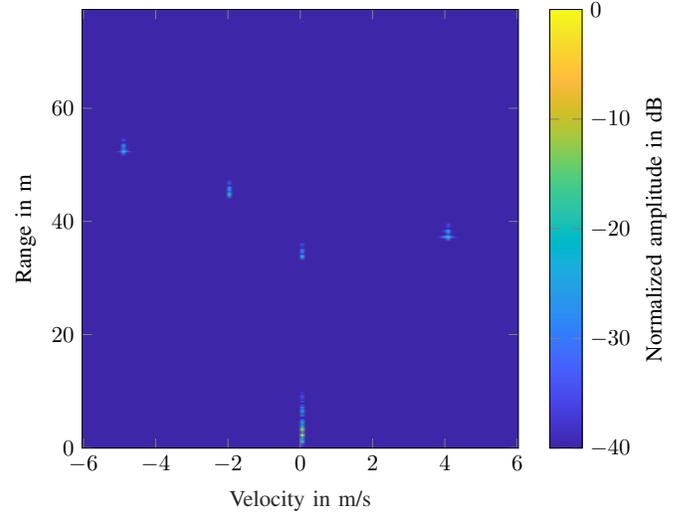}
	\caption{Range-Doppler map of dynamic multi-target measurement}
	\label{fig_meas_range_doppler}
\end{figure}

\begin{figure}[!t]
	\centering
	\includegraphics[width=1\columnwidth]{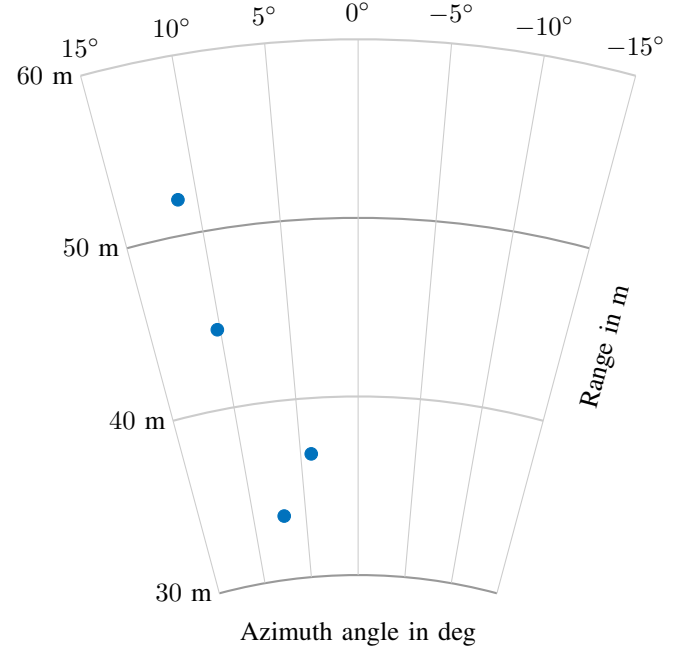}
	\caption{Range angle detections of dynamic multi-target measurement}
	\label{fig_meas_range_angle}
\end{figure}

Finally, a measurement was performed with multiple targets, some of which were subject to a Doppler shift. Four targets with characteristics according to Table \ref{tab_targets} were generated by the RTS simultaneously. Fig. \ref{fig_meas_range_doppler} illustrates the range-Doppler plot measured with a chirp time $T = \SI{41.33}{\micro\second}$ and with $N_\s{chirp} = 120$ chirps. It can be observed, that the targets are generated with the correct range and velocity features. The spurious peaks that occur with an additional range offset to the intended targets are caused by the mismatch of the characteristic impedance between the front ends and the coaxial cables connected to them. The radar signal travels back and forth between the back and front end of the RTS, creating ghost targets with range offsets of multiples of the cable length. The static reflection in close vicinity to the radar can be assigned to the mechanical structure setup of the RTS.

Fig. \ref{fig_meas_range_angle} depicts the detected targets in a range-angle map of the same measurement. All targets are detected at the intended angle, demonstrating the suitability of the approach in the presence of Doppler shifts and underpinning its usability for scenarios with multiple targets.


\section{Conclusion}

The proposed approach enables radar target simulators to generate virtual radar targets at an arbitrary angle of arrival. The mathematical analysis of the signal model presented reveals the constraints that have to be met for a successful steering of the simulated angle. A calibration method to fulfill these requirements was developed and simulatively substantiated. The approach was implemented on a digital radar target s	imulator and the measurement campaign conducted verifies the practical operation of the calibration procedure and investigates the achievable linearity of the adjustable angle of arrival. Furthermore, it demonstrates the suitability of the approach for multi target scenarios comprising of non static targets.



\section*{Acknowledgment}

The authors would like to thank PKTEC GmbH for providing the front end transceiver hardware and Texas~Instruments~Inc. for supplying the radar under test.

\ifCLASSOPTIONcaptionsoff
  \newpage
\fi



%
%
%

\bibliographystyle{IEEEtran}
\bibliography{literature}

%

\begin{IEEEbiography}[{\includegraphics[width=1in,height=1.25in,clip,keepaspectratio]{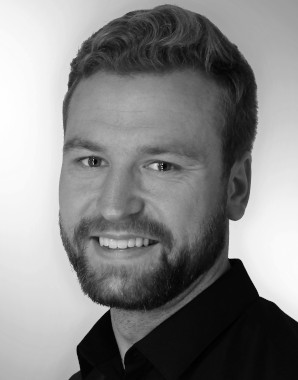}}]{Axel Diewald}
	(GS'18) received the B.Sc. and M.Sc. degrees in electrical engineering and information technology from Karlsruhe Institute of Technology (KIT) in Karlsruhe, Germany, in 2015 and 2017 respectively. Since 2018, he is pursuing a doctorate (Ph.D.E.E.) degree at the Institute of Radio Frequency Engineering and Electronics (IHE) at the KIT. His main fields of research interest are digital radar target simulation for the purpose of automotive radar sensor validation as well as realistic target modeling.
\end{IEEEbiography}

\begin{IEEEbiography}[{\includegraphics[width=1in,height=1.25in,clip,keepaspectratio]{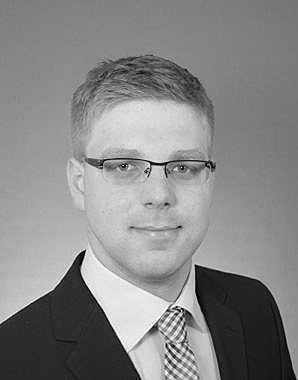}}]{Benjamin Nuss}
	(GS'16) received the B.Sc. and M.Sc. degrees in electrical engineering and information technology from the Karlsruhe Institute of Technology (KIT), Karlsruhe, Germany, in 2012 and 2015, respectively, where he is currently pursuing the Ph.D. degree in electrical engineering at the Institute of Radio Frequency Engineering and Electronics (IHE). His current research interests include orthogonal frequency-division multiplexing-based multiple-input multiple-output radar systems for future automotive applications and drone detection. The focus of his work is the development of efficient future waveforms and interference mitigation techniques for multiuser scenarios.
\end{IEEEbiography}

\begin{IEEEbiography}[{\includegraphics[width=1in,height=1.25in,clip,keepaspectratio]{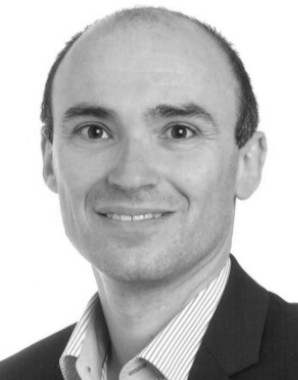}}]{Mario Pauli}	
	(S’04–M’10–SM’19) received the Dipl.Ing. (M.S.E.E.) degree in electrical engineering and Dr.-Ing. (Ph.D.E.E.) from the University of Karlsruhe, Karlsruhe, Germany, in 2003 and 2011, respectively.
	
	Since 2011 he is with the Institute of Radio Frequency Engineering and Electronics at the Karlsruhe Institute of Technology (KIT) as a Senior Researcher and Lecturer. He served as a Lecturer for radar and smart antennas of the Carl Cranz Series for Scientific Education. He is co-founder and the Managing Director of the PKTEC GmbH. His current research interests include radar and sensor systems, RCS measurements, antennas and millimeter-wave packaging.
\end{IEEEbiography}

\begin{IEEEbiography}[{\includegraphics[width=1in,height=1.25in,clip,keepaspectratio]{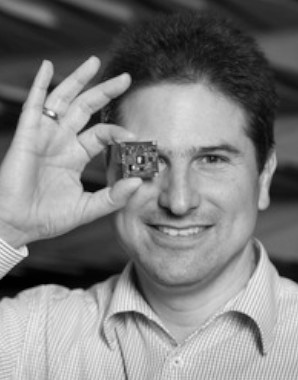}}]{Thomas Zwick}
	(S’95–M’00–SM’06–F’18) received the Dipl.-Ing. (M.S.E.E.) and the Dr.-Ing. (Ph.D.E.E.) degrees from the Universität Karlsruhe (TH), Germany, in 1994 and 1999, respectively. From 1994 to 2001 he was a research assistant at the Institut für Höchstfrequenztechnik und Elektronik (IHE) at the Universität Karlsruhe (TH), Germany. In February 2001, he joined IBM as research staff member at the IBM T. J. Watson Research Center, Yorktown Heights, NY, USA. From October 2004 to September 2007, Thomas Zwick was with Siemens AG, Lindau, Germany. During this period, he managed the RF development team for automotive radars. In October 2007, he became a full professor at the Karlsruhe Institute of Technology (KIT), Germany. He is the director of the Institute of Radio Frequency Engineering and Electronics (IHE) at the KIT. 
	
	Thomas Zwick is co-editor of 3 books, author or co-author of 120 journal papers, over 400 contributions at international conferences and 15 granted patents. His research topics include wave propagation, stochastic channel modeling, channel measurement techniques, material measurements, microwave techniques, millimeter wave antenna design, wireless communication and radar system design.
	
	Thomas Zwick’s research team received over 10 best paper awards on international conferences. He served on the technical program committees (TPC) of several scientific conferences. In 2013 Dr. Zwick was general chair of the international Workshop on Antenna Technology (iWAT 2013) in Karlsruhe and in 2015 of the IEEE MTT-S International Conference on Microwaves for Intelligent Mobility (ICMIM) in Heidelberg. He also was TPC chair of the European Microwave Conference (EuMC) 2013 and General TPC Chair of the European Microwave Week (EuMW) 2017. In 2023 he will be General Chair of EuMW in Berlin. From 2008 until 2015 he has been president of the Institute for Microwaves and Antennas (IMA). T. Zwick became selected as a distinguished IEEE microwave lecturer for the 2013 to 2015 period with his lecture on “QFN Based Packaging Concepts for Millimeter-Wave Transceivers”. Since 2017 he is member of the Heidelberg Academy of Sciences and Humanities. In 2018 Thomas Zwick became appointed IEEE Fellow. In 2019 he became the Editor in Chief of the IEEE Microwave and Wireless Components Letters. Since 2019 he is a member of acatech (German National Academy of Science and Engineering).	
\end{IEEEbiography}

%
%




\end{document}